\documentclass[prb,twocolumn,aps,showpacs]{revtex4}
\usepackage{graphicx}
\usepackage{color,latexsym}
\def\be{\begin{equation}}
\def\ee{\end{equation}}

\begin{document}

\title{Scaling at chiral quantum critical points in two dimensions}
\author{L. Schweitzer$^1$ and P. Marko\v{s}$^2$}
\affiliation{$^1$Physikalisch-Technische Bundesanstalt (PTB), Bundesallee
  100, 38116 Braunschweig, Germany\\
$^2$Institute of Nuclear and Physical Engineering, FEI,
  Slovak University of Technology, 812\,19 Bratislava, Slovakia}

\begin{abstract}
We study the localization properties of electrons moving on
two-dimensional bi-partite lattices in the presence of disorder. The
models investigated exhibit a chiral symmetry and belong to the chiral 
orthogonal (chO), chiral symplectic (chS) or chiral unitary (chU)
symmetry class. The disorder is introduced via real
random hopping terms for chO and chS, while complex random phases 
generate the disorder in the chiral unitary model. In the latter case 
an additional spatially constant, perpendicular magnetic field is also
applied. Using a transfer-matrix-method, we numerically calculate the
smallest Lyapunov exponents that are related to the localization
length of the electronic eigenstates. From a finite-size scaling
analysis, the logarithmic divergence of the localization length at the
quantum critical point at $E=0$ is obtained. We always find for the
critical exponent $\kappa$, which governs the energy dependence of the
divergence, a value close to 2/3. This result differs from the
exponent $\kappa=1/2$ found previously for a chiral unitary model in
the absence of a constant magnetic field. We argue that a strong
constant magnetic field changes the exponent $\kappa$ within the
chiral unitary symmetry class by effectively restoring particle-hole
symmetry even though a magnetic field induced transition from the
ballistic to the diffusive regime cannot be fully excluded.  
\end{abstract}


\maketitle

\section{Introduction}
The properties of electrons moving in disordered two-dimensional (2D)
lattices featuring special symmetries represent a long-standing
challenge for theories of Anderson localization.\cite{EM08} 
It is known that the basic mechanism for electron localization, i.e., 
quantum interference of an electron moving along different paths in
disordered media, depends on the symmetry and topology of the
particular model.\cite{Weg79,Pru84} Recently, the discovery 
of topological insulators\cite{QZ11} and the 
utilization of graphene\cite{Nea05,ZTSK05}---a two-dimensional
honeycomb lattice of carbon atoms---has been producing interest in
this topic anew. In the presence of disorder that preserves the
sublattice symmetry of the underlying two equivalent interpenetrating
triangular sublattices, the energy eigenvalues still come in pairs
$\pm \epsilon_i$ around the Dirac point at $E=0$. The phenomenon of a
quantum critical point at zero energy is not linked to graphene's
hexagonal lattice but can arise also in other bi-partite lattices like
the simple square lattice. 
Disordered systems possessing this property belong to the chiral
symmetry classes.\cite{VZ93,AZ97} One distinguishes three chiral Gaussian
random matrix ensembles: the chiral orthogonal (ChO), the chiral unitary
(ChU), and the chiral symplectic (ChS) one, which differ from each other
by the presence (ChO) or absence (ChU) of time reversal symmetry. The
chS ensemble has time reversal symmetry but spin-rotational symmetry
is lacking, due to spin-orbit interactions in most cases. An additional 
particle-hole symmetry is obeyed in the chO and chS classes.   

In disordered 2D chiral systems, all electronic states are localized
with an energy dependent localization length that diverges at the band
center $E=0$.\cite{MW96,BMSA98,Fur99,ASO03,BC03} The predicted
divergence of the disorder averaged density of states (DOS) obtained
from non-linear sigma-model field
theories\cite{GW91,Gad93,LFSG94,AS99a,BMF00} was apparently observed
in numerical studies,\cite{EK03} although the precise value of the
critical exponent remained unclear.  
On the contrary, in numerical investigations\cite{MDH02} of Motrunich
\textit{et al.\/} for chiral orthogonal systems (time-reversal
symmetry is conserved), the exponent $\kappa$ of the asymptotic energy 
dependence of the disorder averaged DOS, $\rho(E)\sim
E^{-1}\exp(-c|\ln E|^{\kappa})$ was found to be compatible with
$\kappa=2/3$, where $c$ is a nonuniversal constant. 
This result is at variance with the predicted universal exponent
$\kappa=1/2$,\cite{GW91,Gad93,LFSG94,AS99a,FC00,BMF00} that is agreed
to be correct to all orders of perturbation theory. At first, this
difference was believed to be due to the strong disorder limit applied
in Ref.~(\onlinecite{MDH02}). The same result $\kappa=2/3$ was obtained,
however, also in the limit of weak disorder\cite{MRF03} for a
model of non-interacting particles moving on a square lattice via
random nearest neighbor hopping in the presence of a $\pi$-flux
phase.\cite{HWK97} Very recently, it has been pointed out that
non-perturbative effects related to topologically non-trivial
excitations have to be taken into account in field-theoretical
investigations of quantum phase transitions in disordered systems
belonging to the chiral symmetry classes.\cite{KOPM12}     

Besides the divergence of the density of states discussed above, a
corresponding divergence of the localization length with $\kappa=1/2$, 
also predicted by theory\cite{FC00} to occur in a very narrow energy
interval close to $E=0$, has been successfully observed for a chU
model in a numerical study of the smallest Lyapunov exponent only
recently,\cite{MS10} despite several previous
attempts.\cite{Cer00,ERS01,Cer01,ER04,MS07}   
Scaling properties of 2D chiral systems can be studied numerically by
the analysis of the length and disorder or energy dependence of Lyapunov
exponents defined for quasi-one-dimensional (q1D) systems. In the
numerical analysis, strips of finite width $L$ and length $L_z\gg L$ are
calculated by the transfer matrix method. The quantity of interest is the
smallest eigenvalue $x_1$ of the matrix $\ln (T^\dag T)$, where $T$ is
the transfer matrix. Following the finite size scaling method, from
which the localization length of the infinite 2D system can be inferred
by studying q1D samples, we assume that the smallest positive Lyapunov
exponent $\gamma =\lim_{L_z\to\infty} x_1 L/L_z$  is a function of the
ratio $\ln L/\ln\xi$, where $\xi(E)$ is the correlation length. 

Using a bricklayer lattice model with uncorrelated
ran\-dom-magnetic-flux (RMF) 
disorder,\cite{SM08a} distributed uniformly with zero mean, 
a scaling of the dimensionless quantity $z_1=x_1 L/L_z$ was 
observed,\cite{MS10}
\begin{equation}
z_{1}(E,L) = F_{1}\left(\frac{\ln L}{\ln\xi(E)}\right),
\end{equation}
with the scaling function $F_1$ depending only on the ratio $(\ln
L)/(\ln\xi(E))$. 
This model belongs to the chiral unitary symmetry class (broken
time-reversal symmetry). The same exponent was found also for 
a similar model defined on the square lattice.\cite{MS10} 
This result confirms theoretical predictions that for $|E|\ll E_0$ the
localization length is $\xi(E)=\xi_0 \exp[A(\ln(E_0/|E|))^{\kappa}]$ 
with a universal exponent $\kappa=1/2$.\cite{Gad93,FC00} 
However, the recent suggestion\cite{KOPM12} that non-perturbative
effects have to be taken into account applies also to these
investigations. 

It is the aim of our work to improve the understanding of critical
properties at chiral quantum critical points, and to find possible
reasons for the differing critical exponents obtained in the energy
dependence of the DOS and in the divergence of the localization length. 
Therefore, we present in this paper further comprehensive numerical
results obtained for different disorder  strengths in model systems
exhibiting chiral orthogonal, chiral symplectic, or chiral unitary
symmetry in the presence of a very strong perpendicular magnetic
field. The latter situation is quite intriguing because it is 
well known that in a `normal' two-dimensional (diagonal) disordered
system with broken time-reversal symmetry (unitary) all electronic
states are localized. However, this situation converts into the
quantum Hall case, where current carrying critical states appear in
the disorder broadened Landau bands, if a strong normal magnetic field
is applied. Therefore, the question arises whether in the case of
chiral unitary symmetry the properties in the vicinity of the chiral
critical point at $E=0$ are also influenced by the presence of a
strong magnetic field. 
 
We find in all three cases mentioned above a divergence in the energy
dependence of the localization length at $E=0$ with a critical 
exponent $\kappa\simeq 2/3$ similar to the exponent found in
Refs.~(\onlinecite{MDH02,MRF03}) for the diverging disorder averaged
DOS. This value is, however, in contrast to our previous result for
the chiral unitary case without a constant magnetic field, where
$\kappa\simeq 1/2$ was obtained. We argue that a broken particle-hole
symmetry may be responsible for this difference.

\section{Models}
We consider two dimensional (2D) tight binding models defined on
bi-partite square or bricklayer lattices with periodic boundary
conditions applied in both directions. The latter model has recently
been shown to be well suited for describing the electronic properties
of non-interacting particles on graphene's honeycomb lattice in the
presence of disorder.\cite{SM08a}  
Also, in contrast to the square lattice, the van Hove singularities do
not coincide with the quantum critical point at $E=0$ which
facilitates the observation of the energy dependence of the critical
divergences. 
In the absence of diagonal disorder the Hamiltonian reads
\be
{\cal H}=\sum_{\langle n\ne n'\rangle} t_{nn'} c_n^{\dagger}c_{n'}^{},
\ee
where $c_n^{\dagger}$ and $c_{n}$ are the fermionic creation and
annihilation operators at lattice site $n$, and the sum runs over
nearest neighbors only. The disorder is incorporated in the real
random hopping (RRH) terms  
\be\label{eq-ty}
t_y = t_0\exp \frac{W}{t_0}\varepsilon
\ee
pointing in the transversal direction and connecting every other
pair of atoms (those bonds between the zigzag lines on a hexagonal
lattice) in the bricklayer situation. Here, \{$\varepsilon$\} is a set
of uncorrelated random numbers with box probability distribution,
$|\varepsilon|\le 1/2$. In our investigation, the strength of the
disorder is varied between $W/t_0=2$ and $W/t_0=8$.
The spectral bandwidth $\Delta$ is determined by the sum over the mean
of the hopping terms and the coordination number $Z$, 
$\Delta=(Z/2)(\langle t_x\rangle + \langle t_y\rangle)$, with
$t_x/t_0=1$ and   
\be
\langle t_y\rangle = t_0\int_{-1/2}^{+1/2} d \varepsilon \exp
\left[\frac{W}{t_0}\varepsilon\right] 
= \frac{2t_0^2}{W}\sinh\frac{W}{2t_0}. 
\ee
$\Delta$ increases by a factor of 6 when the disorder strength $W$
is increased from $W/t_0=4$ to $W/t_0=8$.

The same off-diagonal disorder type with $W/t_0=4$ was used in the
chiral symplectic model defined on an ordinary square lattice.
We consider a chiral version of the two-dimensional Ando
model.\cite{And89,MS06} The hopping terms $t_{nn'}$ are now $2\times
2$ matrices   
\be
t_\parallel = t_0\left(
\begin{array}{ll}
r & -s \\
-s &   r
\end{array}
\right), \quad
t_\perp = t_y\left(
\begin{array}{ll}
r & is \\
-is &   r
\end{array}
\right)
\ee
where $r^2+s^2=1$, $s=1/2$, and the disorder in the hopping $t_y$
is given by Eq.~(\ref{eq-ty}). 

The chiral unitary model is studied again on a bricklayer
lattice.\cite{SM08a} The hopping terms in the transversal direction
are defined as 
\be
t_y = t_0e^{i\theta_{x,y;x,y\pm a}}, 
\ee
where the phases
$\theta_{x,y;x,y+a}=\theta_{x+2a,y;x+2a,y+a}-\frac{2\pi e}{h}\Phi_{x,y}$ 
are determined by the total flux threading the plaquette at $(x,y)$
\be
\Phi_{x,y} = \frac{p}{q}\frac{h}{e} +\phi_{x,y},
\ee
where $p/q$ ($p$ and $q$ are mutually prime integers) is the rational
number of magnetic flux quanta $h/e$ per plaquette $2a^2$, and
$B=ph/(2qea^2)$ is the magnetic flux density perpendicular to the
two-dimensional lattice. This differs from the random flux model of 
our previous work\cite{MS10}, where the constant magnetic field part
was absent. The random part originates from the magnetic fluxes
$\phi_{x,y}$, which are uniformly distributed $-f/2\le\phi_{x,y}\le
f/2$ with zero mean and disorder strength $0\le f/(h/e)\le 1$.

\section{Method}\label{sect-method}
We calculate the two smallest dimensionless scaling variables $z_1,
z_2$ for quasi-one-dimensional samples having an even width $L$ in the
range $8\le L/a \le 160$ lattice spacings $a$ using the
transfer-matrix-method.\cite{PMR92} For $E=0$, $z_1=2L/\lambda$ is
related to the electronic localization length $\lambda(E=0,L)$.   
The values of $z_1, z_2$ were obtained with the relative accuracy
$\sqrt{\sigma_1}/z_1\sim \alpha\times 10^{-4}$, where $\alpha=3$ for
small $L$, and $\alpha$ increases to 10 for $L/a=128$ and larger. 
For zero energy, we checked that the difference   
\be
z_1(E=0,L) - z_2(E=0,L)
\ee
is smaller than the numerical uncertainty of our data for all $L$.
For the energy dependence, a narrow interval of $|E|\le 10^{-8}t_0$
around the critical point at $E=0$ was considered. 

While investigating various disorder strengths increasing from
$W/t_0=2$ to $W/t_0=8$ for the chiral orthogonal model, it turned out
that the scaling is more difficult to observe for weak disorder. The
reason is that in this case the smallest scaling variable $z_1$ is
less than 1, which means that the related localization length
exceeds the sample width. 
Therefore, putting, for instance,  $W/t_0=2$, we could not observe the 
insulating regime for non-zero energies even on length scales $L/a\le
160$, and our data suffer from strong finite-size effects. Therefore,
in what follows we discuss only data for stronger disorder $W/t_0\ge
5$ with $z_1\ge 1$.  

We fit our numerical data to the polynomial functions
\be\label{xx}
Z_{1,2}(E,L) = z_{1,2}(E=0,L) \pm \sum_n^{n_f} b_n\chi^n 
\ee
where $z_1(E=0,L)$ and $z_2(E=0,L)$ are the numerically obtained
values related to the two smallest Lyapunov exponents at zero
energy. The dimensionless scaling parameter $\chi$ is defined as 
\be\label{eq-chi}
\chi = \frac{[\ln (L)/A]^{x_3}}{[\ln (E_0/|E|)]^{x_2}}.
\ee
We assume the following energy dependence of the correlation length
\be
\ln(\xi(E)/\xi_0) = A |\ln(E_0/|E|) |^{\kappa},
\ee
with unknown parameters $E_0$, $\xi_0$, and critical exponent $\kappa =
x_2/x_3$. 

The total number of fitting parameters is $N_p=4+n_f$, where $n_f$
determines the highest order of polynomial in Eq.~(\ref{xx}). The
additional four parameters are the two exponents $x_2$ and $x_3$, the 
energy $E_0$, and the length scale $\xi_0$. The latter was fixed to
$\xi_0=1$ in all our analysis.\footnote{The analysis shows that the
  parameter $\xi_0$ is not independent of other fitting
  parameters. Changes of $\xi_0$ are compensated by corresponding
  changes of other parameters, but do not influence the value of the
  exponent $\kappa$.} 
In the fitting procedure, we minimize the following function 
\begin{eqnarray}
\label{eq-F}
\lefteqn{F = \frac{1}{N}\sum_n^N \Big(
 \frac{[z_1^{(n)}(E,L) - Z_1(E,L)]^2}{\sigma_1(E,L)}}\\ \nonumber 
 & &{} + \frac{[z_2^{(n)}(E,L) - Z_2(E,L)]^2}{\sigma_2(E,L)}\Big). 
\end{eqnarray}
In Eq.~(\ref{eq-F}), $z_1^{(n)}$ and $z_2^{(n)}$ are the numerically
obtained data for the first and the second smallest scaling variable,
and $\sigma_1$ and $\sigma_2$ are their numerical uncertainties.
To estimate the accuracy of the fit, we create a statistical ensemble
of up to $N_{\rm stat}=50$ sets with initial conditions taken as $z_1
+ \sqrt{\sigma_1}\times \epsilon$ with random number $|\epsilon|<1$.

After fitting the obtained numerical data for $z_1$ and $z_2$ to the
function (\ref{xx}), we change the number of fitting parameters (i.e.,
the order $n_f$ of the polynomial) in order to check the reliability
of the fit. Although we do not expect strong finite size effects, we 
also vary the smallest size of the systems taken into account in the
data ensembles. Another check of the stability of the resulting
exponent is to reduce the energy interval, for instance to values
$E<10^{-13}t_0$. 

\begin{table}
\caption{The critical parameters for models with RRH or RMF disorder
  with a strong constant magnetic field as obtained by the fitting
  procedure described in Section \ref{sect-method}. In the ensembles
  marked by $^*$, a restricted energy interval $|E|<10^{-13}t_0$ was
  used instead of $|E|<10^{-8}t_0$. The labeling of the chiral
  orthogonal chO[\ ], symplectic chS[\ ] and unitary models chU[\ ] is
  explained in the text below. These models exhibit a critical
  exponent close to 2/3. 
}
\medskip
\begin{small}
\noindent%
\begin{tabular}{|l|l|l|l|l|l|}
\hline
$N_p$  & $L/a$ &   $\ln E_0$   &   $\kappa$   &  $x_3$   &  $\frac{F_{\rm min}}{2}$ \\
\hline
\hline
\multicolumn{6}{|l|}{chO: RRH $W/t_0=5$, $z_1=1.35$} \\
\hline
5    &  12-96 &$ -1.60\pm 0.41$  &    $0.813\pm 0.013$  &$  1.100\pm 0.03$ & 0.29 \\
$5^*$    &  12-96 &$ -2.3\pm 1.5$  &    $0.801\pm 0.03$  &$  1.17\pm 0.02$ & 0.11  \\
\hline
\multicolumn{6}{|l|}{chO: RRH $W/t_0=6$, $z_1=1.79$} \\
\hline
5    &  12-160 &$ -5.50\pm 0.20$  &    $0.665\pm 0.006$  &$  1.220\pm 0.003$ & 2.24 \\
$5^*$    &  24-160 &$ -4.17\pm 2.02$  &    $0.724\pm 0.046$  &$  1.160\pm 0.11$ & 0.28 \\
\hline
\multicolumn{6}{|l|}{chO: RRH $W/t_0=7$, $z_1=2.23$} \\
\hline
5    &  12-128 &$ -3.65\pm 0.18$  &    $0.675\pm 0.005$  &$  1.272\pm 0.002$ & 4.89 \\
$5^*$    &  12-128 &$ -4.4\pm 1.5$  &    $0.693\pm 0.03$  &$  1.204\pm 0.02$ & 0.39  \\
\hline
\multicolumn{6}{|l|}{chO: RRH $W/t_0=8$, $z_1=2.65$} \\
\hline
$5^*$    &  12-96 &$ -6.4\pm 0.5$  &  $0.628\pm 0.01$  &$  1.25\pm 0.03$ & 1.71  \\
\hline
\hline
\multicolumn{6}{|l|}{chS: RRH $W/t_0=4$, $z_1=1.97$} \\
\hline
$4$    &  8-160 &$ -5.4\pm 0.1$  &    $0.687\pm 0.049$  &$  1.403\pm 0.021$ & 7.40  \\
$4^*$    &  8-160 &$ -7.5\pm 0.5$  &    $0.692\pm 0.015$  &$  1.257\pm 0.004$ & 1.22  \\
$4$    &  16-160 &$ -6.1\pm 0.1$  &    $0.676\pm 0.058$  &$  1.409\pm 0.041$ & 6.40  \\
$5$    &  16-160 &$ -3.4\pm 0.1$  &    $0.796\pm 0.078$  &$  1.045\pm 0.033$ & 2.83  \\
$5$    &  48-160 &$ -6.1\pm 0.7$  &    $0.664\pm 0.028$  &$  1.186\pm 0.025$ & 0.98  \\
\hline
\hline
\multicolumn{6}{|l|}{chU[1/16, 5.0]: $B=1/16\,h/(2ea^2)$, RRH $W/t_0=5$,} \\
\multicolumn{6}{|l|}{\hfill $z_1=2.47$}\\
\hline
6    &  32-96 &$ -1.74\pm 0.37$  &    $0.636\pm 0.01$  &$  1.27\pm 0.13$ & 1.4 \\
7    &  32-96 &$ -1.23\pm 0.64$  &    $0.648\pm 0.02$  &$  1.17\pm 0.17$ & 0.8 \\
\hline
\multicolumn{6}{|l|}{chU[1/12, 0.5]: $B=1/12\,h/(2ea^2)$, RMF $f=0.5\,h/e$,}\\ 
\multicolumn{6}{|l|}{\hfill $z_1=1.03$} \\
\hline
4    &  24-72 &$ -6.74\pm 0.24$  &    $0.567\pm 0.008$  &$  1.75\pm 0.003$ & 3.504 \\
5    &  24-72 &$ -4.14\pm 0.65$  &    $0.643\pm 0.02$  & $ 1.313\pm 0.02 $& 2.142 \\
6    &  24-72 &$ -3.45\pm 0.65$  &    $0.663\pm 0.02$  & $ 1.332\pm 0.23 $& 2.055 \\
7    &  24-72 &$ -3.02\pm 0.96$  &    $0.676\pm 0.03$  & $ 1.411\pm 0.33 $& 2.016 \\
\hline
\end{tabular}
\end{small}
\label{tabulka}
\end{table}

\begin{table}
\caption{The critical parameters for models with RMF disorder and zero 
  or small constant magnetic fields as obtained by the fitting
  procedure described in Section \ref{sect-method}. A critical
  exponent close to 1/2 is found. The energy interval used was
  $|E|<10^{-10}t_0$. The labeling of the various 
  models chU[\ ] is explained in the text.   
}
\medskip
\begin{small}
\noindent%
\begin{tabular}{|l|l|l|l|l|l|}
\hline
$N_p$  & $L/a$ &   $\ln E_0$   &   $\kappa$   &  $x_3$   &  $\frac{F_{\rm min}}{2}$ \\
\hline
\hline
\multicolumn{6}{|l|}{chU[0, 0.5]: $B=0$, RMF $f=0.5\,h/e$, $z_1=z1 = 1.555$} \\
\hline
6  & 8-160  & $-2.6\pm 1.6$  & $0.414\pm 0.04 $  & $2.49\pm 0.44$ & 0.91\\  
7  & 8-160  & $0.25\pm 2.5$  & $0.476\pm 0.05 $  & $1.98\pm 0.5 $ & 1.43 \\
8  & 8-160  & $1.86\pm 1.1$  & $0.511\pm 0.025$  & $2.01\pm 0.44$ & 1.28 \\
\hline
\multicolumn{6}{|l|}{chU[0, 0.7]: $B=0$, RMF $f=0.7\,h/e$, $z_1=1.22$} \\
\hline
$5$    &  24-96 & $-4.5 \pm 1.5 $  & $0.515 \pm 0.04$  & $1.86 \pm 0.20$ & 2.43  \\
$5$    &  32-96 & $-3.5 \pm 1.5 $  & $0.55 \pm 0.035$  & $1.68 \pm 0.11$ & 2.01  \\
$5$    &  40-96 & $-4.0 \pm 1.7 $  & $0.54 \pm  0.04$  & $1.65 \pm 0.10$ & 1.86  \\
\hline
\multicolumn{6}{|l|}{chU[0, as]: $B=0$, RMF $f/(h/e)=-0.25, 0.125$, $z_1=1.328$}  \\
\hline
5  & 16-96 & $-4.58\pm 1.8$ & $0.474\pm 0.04$ & $2.156\pm 0.02$ & 1.36\\
6  & 16-96 & $-1.66\pm 2.6$ & $0.535\pm 0.05$ & $2.40\pm 0.38$  &1.22\\
\hline
\hline
\multicolumn{6}{|l|}{chU[1/10K, 0.5]: $B=1/10000\,h/(2ea^2)$, RMF $f=0.5\,h/e$,}\\ 
\multicolumn{6}{|l|}{\hfill $z_1=1.5559$} \\
\hline
7    & 24-96  &   $0.38\pm 4$ & $0.464\pm 0.07$ &  $2.38\pm 1.14$ & 2.01 \\
8    & 24-96  &   $1\pm 1$    & $0.475\pm 0.02$ &  $3.16\pm 0.75$ & 2.39\\
\hline
\multicolumn{6}{|l|}{chU[1/1K, 0.5]: $B=1/1000\,h/(2ea^2)$, RMF $f=0.5\,h/e$,}\\ 
\multicolumn{6}{|l|}{\hfill $z_1=1.557$} \\
\hline
6    & 24-96 &  $-0.7\pm 3.7$  & $0.448\pm 0.063$ & $2.27\pm 0.64$   &   1.82\\ 
7    & 24-96 &  $2.7\pm  2.5$  & $0.511\pm 0.043$ & $2.40\pm 0.75$   &   2.52\\
8    & 24-96 &  $2.8\pm 1.4 $  & $0.512\pm 0.023$ & $2.73\pm 0.44$   &   2.22\\
\hline
\end{tabular}
\end{small}
\label{tabulka2}
\end{table}


\section{Results}
The results for all models analyzed are summarized in Table~\ref{tabulka}.
For the chiral orthogonal model, we present data for four different
values of the disorder strength in the range $5\le W/t_0\le 8$. These
disorders are strong enough to assure that the smallest scaling variable
is $z_1>1$, so that we are in the localized regime for non-zero
energies. The exponents $\kappa$ are close to the value 2/3. Small
deviations are seen for weak disorder $W=5$. 

For the chiral symplectic system, we analyzed only one value of the
disorder strength, $W/t_0=4$. The obtained value for the critical
exponent is compatible with $\kappa=2/3$ and agrees well with the one 
found for the chiral orthogonal symmetry.  

For chiral unitary systems, we consider a model with a constant
magnetic flux density $B=(1/12)\,h/(2ea^2)$ and RMF disorder strength
$f/(h/e)=0.5$ (labeled  chU[1/12, 0.5]), and also a model where the
constant magnetic field $B=(1/16)\,h/(2ea^2)$ is combined with real
hopping terms $t_y$ as given by Eq.~(\ref{eq-ty}) (model chU[1/16,
 5.0]). Both systems exhibit a critical exponent consistent with a
value 2/3. Please note that the energies of the quantum Hall critical
states belonging to the zeroth Landau band are split\cite{SM08a} by
the disorder, and so they do not interfere with the chiral critical
point at $E=0$ studied here.   

\begin{figure}
\includegraphics[width=8.0cm,clip]{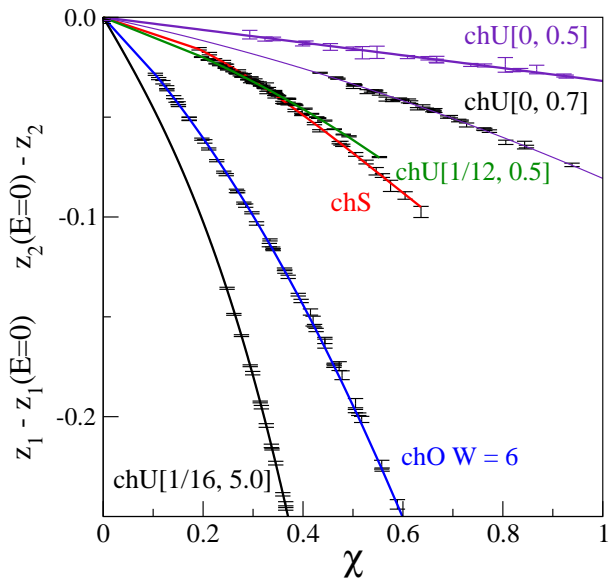}
\caption{(Color online) The difference $-|z_{i}(E,L)-z_{i}(E=0,L)|$ of
  the two smallest dimensionless scaling variables plotted as a
  function of the parameter $\chi$ with corresponding fits (solid
  lines) for some systems listed in Table~\ref{tabulka} and
  \ref{tabulka2}. The previously obtained chU[0, 0.5] data\cite{MS10}
  are shown only in part since $\chi$ increases up to values of $\sim
  10$ in this case.   
}
\label{beta-all}
\end{figure}

To compare the new results with our previous data published in
Ref.~\onlinecite{MS10} for the chiral unitary system without a
constant magnetic field, we use the present fitting method also for
the analysis of our previous data. In Table~\ref{tabulka2} we see 
that the critical exponent for the model chU[0, 0.5] is consistent
with the value 1/2 as reported previously,\cite{MS10} but differs 
significantly from the value 2/3 obtained here for all other
ensembles. Also, in order to check a possible disorder dependence of
$\kappa$, again a bricklayer model with zero constant magnetic field
but with a stronger RMF disorder strength $f=0.7\,h/e$ (labeled
chU[0, 0.7]) was calculated. We obtain the same value $\kappa\simeq
1/2$ and so confirming our previous results.    
In Fig.~\ref{beta-all} we compare
some of the raw numerical data for $z_i(E,L)-z_i(E=0,L)$, $i=1,2$, with
the fit given by Eq.~(\ref{xx}). 
Some other models, listed in Table~\ref{tabulka2}, which also exhibit
an exponent $\kappa=1/2$ will be discussed in the next section.

\section{Discussion and Conclusions}
Our numerical data for two-dimensional models with various chiral
symmetry confirm the logarithmic energy dependence of the localization
length reported previously.\cite{MS10} Here, we find that this
logarithmic dependence is determined by the critical exponent
$\kappa$, which is universal and close to 2/3 for chO, chS, and chU
(with magnetic field) symmetry classes. The only exception is the
chiral unitary system with zero constant magnetic field where
$\kappa\simeq 1/2$. We have confirmed this previous result by
additional calculations for a RMF disorder strength $f=0.7\,h/e$. 
Therefore, the chiral unitary models with and without constant
magnetic field studied here exhibit different critical behaviors. 

One source for this outcome could be that the distribution of the
random magnetic fluxes, which was chosen to be symmetric about zero
flux in the $B=0$ situation, turns into an asymmetric one due to the
addition of a spatially constant magnetic flux in the finite magnetic
field case. We have checked that an asymmetric distribution of
RMF-fluxes still leads to a critical exponent 1/2 as long as the
average magnetic flux remains zero. In these calculations, the random
flux was taken to be $-0.25\,h/e$ with probability 1/3 and $0.125\,h/e$
with probability 2/3 so that the distribution of random fluxes is
asymmetric albeit the mean value is zero. Using a 6-parametric fit
as given by Eq.~(10), we obtained that $\kappa = 0.535 \pm 0.05$ for
$16\le L/a \le 96$. The details for the antisymmetric RMF distribution 
chU[0, as] can be seen in Table~\ref{tabulka2}. 

\begin{figure}[t]
\includegraphics[width=7.8cm]{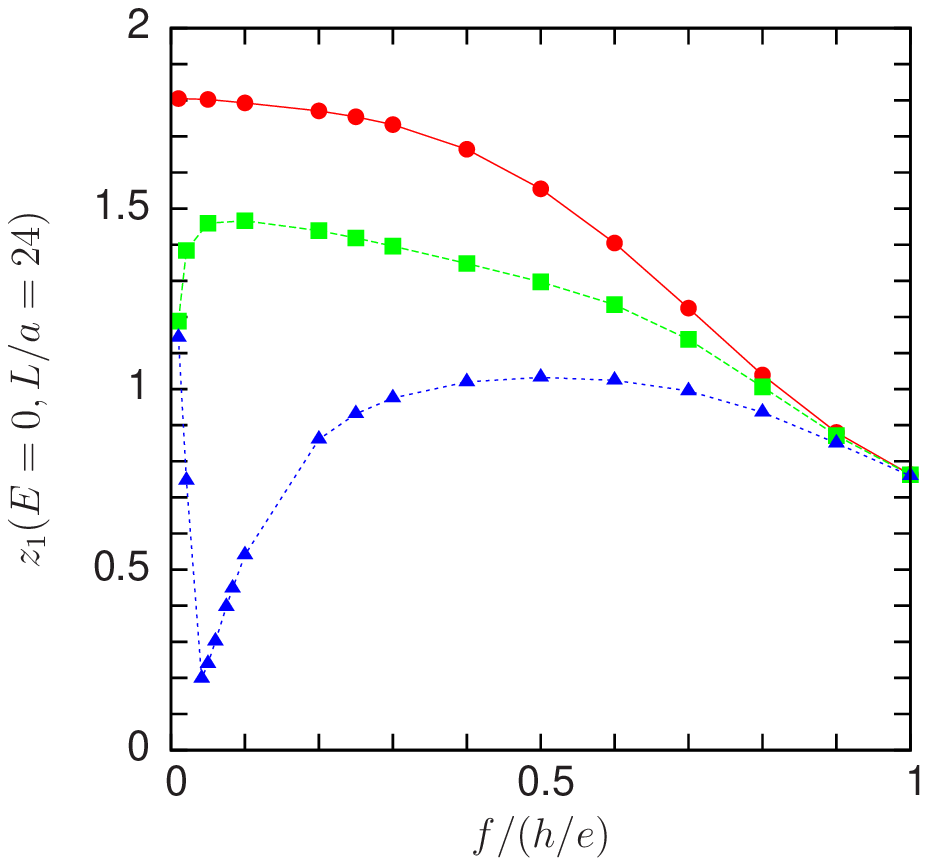}
\caption{(Color online) The smallest scaling variable $z_1(E=0,f)$
  versus RMF 
  disorder strength $f$. The constant magnetic field is zero (red
  dots), $B=1/24\times h/(2ea^2)$ (green squares), and $B=1/12\times
  h/(2ea^2)$ (blue triangles), respectively. These results do not
  dependent on the system width in the range $24\le L/a \le
  128$ checked.   
}
\label{z1f}
\end{figure}

Another reason for the different exponents
could be that, for the largest system size studied, 
the disorder strength is still too small in the zero constant magnetic
field situation. Due to the periodicity, the effect of the random flux
disorder $-f/2\le\phi_{x,y}\le f/2$ appearing in the phases
$\exp(-i\theta_{x,y+a;x,y})$ with
$\theta_{x,y+a;x,y}=\theta_{x+2a,y;x+2a,y+a}-2\pi e \phi_{x,y}/h$     
is bounded, being maximal for disorder strength $f=1.0\,h/e$ for the
bricklayer lattice. 
The dependence of the smallest scaling variable
$z_1(E=0,f)$ on the disorder strength $f\ge 0.01$ is shown in
Fig.~(\ref{z1f}) for $B=0$, $p/q=1/24$, and $p/q=1/12$, respectively. 
In the small disorder regime, the behavior is complicated and not
important for the problem of different critical exponents discussed
here. For large $f$, the $z_1(E=0,f)$ decrease in all RMF models,
independent of the presence or absence of a constant magnetic field,
and become equal $\sim 0.76$ for $f=h/e$. Usually, a decreasing
$z_1$ is equivalent to an increasing localization length 
$\lambda = 2L/z_1$. A similar increase of the
two-terminal conductance with increasing RMF-disorder strength has
been reported previously.\cite{SM08} The behavior of the RMF-models
is opposite to the chO and chS situation where $z_1(E=0,W)$ becomes
larger with increasing disorder strength $W$. However, when comparing
the various values of $z_1(E=0)$  obtained for different models (see
Table~\ref{tabulka}), which are related to the effective disorder
strength, the above suggestion that too weak a disorder was
applied in the RMF-model without $B$-field seems to be unlikely, and
we do not believe that this is the correct explanation for the
observed difference in the critical exponents.  

Another distinction between the two chU cases is the behavior of the
density of states $\rho(E)$ very close to $E=0$. Although both
the chU models with and without strong magnetic field exhibit a narrow
depression,\cite{Sch09} with $\rho(E)$ going to zero at the critical
point, the disorder dependence is quite different. With increasing 
random-magnetic-flux disorder strength, the energy range of the
depression gets narrower in the absence of a constant $B$ while it
increases when a strong perpendicular magnetic field is present. The
appearance of such `microgaps' has been attributed to the
non-perturbative ergodic regime\cite{AS99a,SA01,AM01} of the chiral
model, i.e., the localization length greatly exceeds the sample size
so that for time scales large compared to the diffusion time the
particle extends over the whole sample. A similar behavior was
observed also for the chS model.\cite{MS12} We note, however, that DOS
depressions (`microgaps') are absent in our numerical calculations
performed for the chO case. For strong disorder, a narrow peak appears
instead at $E=0$,\cite{MS12} presumably originating from the different
type of disorder (real random hopping) that in contrast to the complex
random phases in the RMF models can lead to isolated sites having
eigenenergies around $E=0$. This extra peak is observed for both the
bricklayer and the square lattice systems. Nevertheless, it is unclear
whether the different disorder dependence of the DOS around the
critical point can account for the observed difference in the critical
exponents.  

Finally, we would like to remind the reader that the presence of a
strong magnetic field causes a special topological term in the
appropriate field theories.\cite{LLP83,Pru84,Wei87} This extra term is 
responsible for the occurrence of current carrying states that are
essential for the explanation of the integer quantum Hall effect, and
so it accounts for the difference between the ordinary Gaussian
unitary ensemble and the quantum Hall situation, i.e., the appearance
of critical electronic states in the latter case. 

Furthermore, the
strong magnetic field may effectively restore the particle-hole
symmetry that is usually lacking when time-reversal symmetry is
broken. The recovery of the particle-hole symmetry can take place when  
the scattering between electronic states belonging to different
disorder broadened Landau bands is suppressed by a large energy
separation. Thus, a single Landau band model containing a chiral 
quantum critical point at $E=0$ is effectively created in the chiral
unitary case with an additional very strong, spatially constant,
perpendicular magnetic field. We therefore suggest that the presence
or absence of particle-hole symmetry is responsible for the different
critical exponents found in our numerical calculations of various
chiral unitary models. To check this view, we investigated also chiral 
unitary models with weaker constant magnetic fields $p/q=1/10000$
and $p/q=1/1000$ (see chU[1/10K, 0.5] and chU[1/1K, 0.5] in
Table~\ref{tabulka2}) and found critical exponents $\simeq 1/2$ as in 
the magnetic field free case.      

In conclusion, we calculated the two smallest Lyapunov exponents of
lattice models belonging to the chiral orthogonal, chiral symplectic,
and to the chiral unitary class in the presence of a strong spatially
constant magnetic field. We found in all cases a critical exponent
$\kappa\simeq2/3$ that governs the divergence of the localization
length at the band center $E=0$. This result is in agreement with
previous calculations for the diverging energy dependence of the
density of states.\cite{MDH02,MRF03} It is, however, at variance with
the earlier analytically obtained results.\cite{Gad93,FC00} While
reasons for this difference were suggested in
Refs.~(\onlinecite{MDH02,MRF03,EM08,KOPM12}), 
the origin of the numerically\cite{MS10} obtained value
$\kappa\simeq1/2$ for chiral unitary lattice models in the absence of
a strong perpendicular magnetic field is still an open question.  
The latter result was again corroborated in the present study
and suggested to be due to the absence of particle-hole symmetry in
the RMF-model without a strong spatially constant magnetic field. 
However, due to the limited achievable system size in the numerical
calculations, a magnetic field induced transition from the ballistic
to the diffusive regime as discussed above, cannot be completely
excluded.

\section*{Acknowledgments} P.~M. thanks Project VEGA 0633/09 for
financial support.

\bibliographystyle{apsrev}

\end{document}